 \documentclass[12pt]{article}
\newcommand{\be}{\begin{equation}}
\newcommand{\ee}{\end{equation}}
\newcommand{\ba}{\begin{eqnarray}}
\newcommand{\ea}{\end{eqnarray}}

\newcommand{\inc}{{\it i}}
\newcommand{\efbold}{\mbox{{\boldmath $\vec f$}}}
\newcommand{\robold}{\mbox{{\boldmath $\vec \rho$}}}
\newcommand{\erbold}{\mbox{{\boldmath $\vec r$}}}
\newcommand{\Phibold}{\mbox{{\boldmath $\vec \Phi$}}}
\newcommand{\mubold}{\mbox{{\boldmath $\vec \mu$}}}
\newcommand{\xbold}{\mbox{{\boldmath $\vec x$}}}
\begin{document}
\date{}
\title{
   ~~~~~~~~~~~~~~~~~~~~~~~~~~~~~~~~~~~~~~~~~~${~}^{astro-ph/0212245}$\\
    ~${~}^{~}$\\
{\Large{\bf Implicit gauge symmetry emerging\\ in the N-body
problem of celestial mechanics.}}\\
}
\author{ {\Large{Michael Efroimsky}}\\
 {\small{Institute for Mathematics and its Applications}}\\
  {\small{University of Minnesota, Minneapolis MA 55455}}\\
{\small{e-mail: efroimsk @ ima.umn.edu~}}\\
 ~\\
  {\small{Present address: ~US Naval Observatory, ~Washington DC 20392}}
}
 \maketitle
 \begin{abstract}
We revisit the Lagrange and Delaunay systems of equations of
celestial mechanics, and point out  a previously neglected aspect
of these equations: in both cases  the orbit resides  on a certain
9(N-1)-dimensional submanifold of the 12(N-1)-dimensional space
spanned by the orbital elements and their time derivatives. We
demonstrate that there exists a vast freedom in choosing  this
submanifold. This freedom of choice (=freedom of gauge fixing)
reveals a symmetry hiding behind Lagrange's and Delaunay's
systems, which is, mathematically, analogous to the gauge
invariance in electrodynamics.  Just like a convenient choice of
gauge simplifies calculations in electrodynamics, so the freedom
of choice of the submanifold may, potentially, be used to create
simpler schemes of orbit integration. On the other hand, the
presence of this feature may be  a previously unrecognised source
of numerical instability.\\
\\

\end{abstract}

 \pagebreak

\section{~~Introduction~~~~~~~~~~~~~~~~~~~~~~~~~~~~~~~~~~~~~~~~~~~~~~~~~~~~~~~~~~~~~~~~~~~~~~~~~~~~~~~~~~~~}

\subsection{The Variation-of-Parameters Method}

The method of variation of parameters (VOP) was invented in the
middle of XVIII-th century by Euler (1748, 1753) as a tool for
treating highly-nonlinear problems emerging in the context of
celestial mechanics. Though Euler himself eventually lost interest
in this approach, his line of research was intensively continued
by Lagrange who employed it for deriving his celebrated system of
equations describing evolution of the orbital parameters (which in
the astronomical parlance are called "orbital elements").

In the modern textbooks, the VOP method is normally introduced as
one of the means by which one can solve an inhomogeneous linear
differential equation: one first finds all solutions of the
appropriate linear homogeneous equation, and then instills time
dependence into the coefficients in the linear combination of
these solutions. Here follows the easiest example:
 \ba
\ddot{x}\;+\;p(t)\;\dot{x}\;+\;q(t)\;x\;=\;F(t)\;\;.
 \label{1.1}
  \ea
 To
solve this inhomogeneous equation, one starts out with the
homogeneous one:
 \be
 \ddot{x}\;+\;p(t)\;\dot{x}\;+\;q(t)\;x\;=\;0\;\;.
 \label{1.2}
  \ee
A linear combination of its two fundamental solutions will read:
$\; S\;x_1(t)\;+\;C\;x_2(t)\;$.
 The recipe has it that at this point one should
look for a solution to (\ref{1.1}) in ansatz
 \be
 x\;=\;S(t)\;x_1(t)\;+\;C(t)\;x_2(t)\;\;.
 \label{1.4}
 \ee
Since the functions $x_{1,2}(t)$ are known, what one has to do is
just to find $S(t)$ and $C(t)$. Equation  (\ref{1.1}) will, by
itself, be insufficient for determining two independent functions.
The excessive freedom can be removed through a by-hand imposure of
an extra equality often chosen as
 \be
 \dot{S}\;x_1\;+\;\dot{C}\;x_2\;=\;0\;\;.
 \label{1.5}
 \ee
It greatly simplifies the expressions for $\dot x$ and $\ddot x$:
 \be \dot
x\;=\;S\;\dot{x}_1\;+\; C\;\dot{x}_2\;\;\;\;,\;\;\;\;\;\;\;\;\;\;
\ddot{x}\;=\;\dot{S}\;\dot{x}_1\;+\;\dot{C}\;\dot{x}_2\;+\;C\;\ddot{x}_1\;+\;
C\;\ddot{x}_2\;\;,
 \label{1.6}
 \ee
substitution whereof in (\ref{1.1}) entails:
 \be
 \dot S\;\dot x_1\;+\;\dot C\;\dot x_2\;=\;F\;\;.
 \label{1.7}
 \ee
Together with (\ref{1.5}), the latter yields:
 \be
 S\;=\;-\;\int^{t}\;\frac{F(t')\,x_2(t')}{W[x_1,\;x_2](t')}\;dt'\;\;\;\;,\;\;\;\;\;\;\;\;\;\;
 C\;=\;\int^{t}\;\frac{F(t')\,x_1(t')}{W[x_1,\;x_2](t')}\;dt'
 \label{1.8}
 \ee
 \be
{\mbox{where}} \;\;\;\;\;\;\;\;\;\;\;\;\;\;\;\;\;\;
W[x_1,\;x_2](t)\;\equiv\;x_1(t)\;\dot{x}_2(t)\;-\;x_2(t)\;\dot{x}_1(t)\;\;.\;\;\;\;\;\;\;\;\;
 \label{1.9}
 \ee
 This traditional way of introducing the method of
variation of parameters (VOP) is pedagogically flawed because it
does not illustrate the full might and generality of this
approach.
What is important is that the initial equation, whose solution(s)
is (are) assumed to be known, does not necessarily need to be
linear. Moreover, the parameters to be varied should not
necessarily be the coefficients in the linear combination of
solutions. Historically, Euler and Lagrange developed this
approach in order to solve the nonlinear equation (\ref{2.9}), so
the parameters to vary (the orbital Keplerian elements) were
{\bf{not}} coefficients of a linear combination of solutions to
the homogeneous equation. Rather, these were the constants of the
unperturbed motion (i.e., the quantities which were conserved in
the homogeneous (2-body) case but no longer conserved in the
inhomogeneous (N-body) case).

A pretty evident aspect of this method is that any choice of the supplementary
condition different from (\ref{1.5}) will render the same solution $\;y(x)\;$,
even though the expressions for the ``constants" $\;C_{1,\,2}\;$ will differ
from (\ref{1.8}). In other words, the solution $\;y(x)\;$ remains invariant
under a certain family of transformations of $\;C_{1,\,2}\;$. This internal
symmetry is inherent in any problem that can be approached through the VOP
method, including the N-body problem of celestial mechanics where the
parameters $\;C_i\;$ are the constants of the unperturbed motion. To
illustrate this, let us consider a trivial example kindly offered to the
author by William Newman.

\subsection{Newman's Example}

As a simplest instance, consider a forced harmonic oscillator
 \ba
 \nonumber H(x,p) = \frac{p^2}{2} + \frac{x^2}{2} + x \,F(t)
 \ea
that leads to the well known initial-condition problem
 \be
 \ddot x
 + x = F(t)\;\;\;,\;\;\;\;\;{\mbox{with}}\;\;x(0)\;\;{\mbox{and}}\;
\;\dot x (0)\;\;{\mbox{known}}.
 \label{5.1}
 \ee
 As prescribed by
the method of variation of parameters, we seek a solution in
ansatz \be x = S(t) \sin t + C(t) \cos t \;\;\;. \label{5.2} \ee
which yields \be \dot x = \left\{\dot S(t) \sin t + \dot C(t) \cos
t \right\}\;+\; S(t) \cos t - C(t) \sin t \;\;\;. \label{5.3} \ee
The standard procedure implies that we put $\;\dot S(t) \sin t +
\dot C(t) \cos t\,=\,0\;$, in order to get rid of the ambiguity.
The by-hand imposure of this equality is convenient but not
necessarily required. Any other way of fixing the ambiguity, like
for example, \be \dot S(t) \sin t + \dot C(t) \cos t \,=\,\Phi(t)
\label{5.4} \ee will be equally good. Then \be \ddot x = \dot
\Phi\;+\;\dot S(t) \cos t - \dot C(t) \sin t - S(t) \sin t - C(t)
\cos t\;\;\;, \label{5.5} \ee whence \be \ddot x + x = \dot
\Phi\;+\;\dot S(t) \cos t - \dot C(t) \sin t \label{5.6} \ee Thus
one faces the system \ba \nonumber
\dot \Phi\;+\;\dot S(t) \cos t - \dot C(t) \sin t =\;F(t)\\
\label{5.7}\\
\nonumber \dot S(t) \sin t + \dot C(t) \cos t \,=\,\Phi(t)\;, \ea
the first line being the equation of motion (obtained through
combining (\ref{5.6}) with (\ref{5.1})), and the second line being
identity (\ref{5.4}).\footnote{This identity becomes a constraint
if one fixes the functional form of function $\;\Phi\;$ which thus
far has remained arbitrary.}  The system trivially resolves to \ba
\nonumber
\dot S\;=\;\,F\;\cos t\;-\;\frac{d}{dt} \,\left( \Phi \cos t \right)\\
\label{5.8}\\
\nonumber \dot C\;=\,-\,F\;\sin t\;+\;\frac{d}{dt} \,\left( \Phi
\sin t \right) \ea The function $\;\Phi\;$ still remains
arbitrary,\footnote{Mind that this arbitrariness of $\;\Phi\;$
stays, no matter what the initial conditions are to be. Indeed,
for fixed $x(0)$ and $\dot x(0)$, the system $\;C(0) = x(0)\;,\;
\; \Phi (0) + S(0) = \dot x(0)\;$ solves for $S(0)$ and $C(0)$ for
an arbitrary choice of $\Phi (0)$.} as can be easily seen either
from the above derivation or from direct substitution of
(\ref{5.8}) in (\ref{5.1}). Integration of (\ref{5.8}) trivially
yields
 \ba
 \nonumber
S\;=\;\,\int^{t}\,F\;\cos t'\;dt'\;-\; \Phi \cos t \;+\;c_1\\
 \label{5.9}\\
\nonumber C\;=\,-\,\int^{t}\,F\;\sin t'\;dt'\;+\;\Phi \sin t
\;+\;c_2
 \ea
inclusion whereof into (\ref{5.2}) entails \ba x=S
\sin t + C \cos t=\,-\,\cos t\,\int^t F\,\sin t'\,dt'\,+\, \sin
t\,\int^t F\,\cos t' \, dt'\,+\,c_1\,\sin t\,+\,c_2\,\cos
t\;\;\;\;\; \label{5.10} \ea with the $\Phi$-terms cancelled out.
Analogy of this simple example with the osculating-element
formalism is evident. Variable $\;x\;$ considered in this example
is analogous to the Cartesian coordinates and velocities in
Section III, while parameters $\;S\;$ and $\;C\;$ are counterparts
of the osculating elements. Equations (\ref{5.8}) are analogues to
(\ref{4.15}) - (\ref{4.20}). These expressions for $\;\dot S\;$
and $\;\dot C\;$ through $\,\Phi\,$ show that it is impossible (at
least, in this particular example) to pick up such $\;\Phi\;$ that
the right-hand sides in both equations (\ref{5.8}) simplify.
Newman's example also thwarts one's hope to separate the time
scales: as can be seen from (\ref{5.9}), even when the
perturbation $\,F(t)\,$ is a slow function of time, the time
evolution of the ``osculating elements'' $\;S\;$ and $\;C\;$ is
determined by the ``fast'' time scale associated with $\;\sin t\;$
and $\;\cos t\;$, i.e., with the solutions for the homogeneous
equation.

The above setting was much simpler than the N-body case where one
is faced with nonlinearity and the parameters to be varied are
{\it{not}} coefficients of a linear combination of solutions to a
homogeneous equations. However, Newman's example well illustrates
the origin of internal symmetry in the VOP context. This symmetry
is a generic feature that shows itself in all implementations of
the VOP approach, including the N-particle dynamics.

Another good thing about Newman's example is that it illustrates the
impossibility of time-scales' separation in the VOP method: in the right-hand
side of (\ref{5.9}) the expressions standing under the integral contain both
the free-oscillator and disturbing frequencies. No matter what gauge
$\;\Phi\;$ one chooses, at least one of these integrals will contain a mixture
of frequencies. This way, if we have a fast oscillator subject to a slow
force, we still cannot reduce its motion to a slow evolution of the amplitudes
$S(t)$ and $C(t)$; instead, the time dependence of these amplitudes will
contain fast terms (which is very counter-intuitive).

 \pagebreak
\section{~~Background~~~~~~~~~~~~~~~~~~~~~~~~~~~~~~~~~~~~~~~~~~~~~~~~~~~~~~~~~~~~~~~~~~~~~~~~~~~~~~~~~~~~~~~~~~~}

Solar-System dynamics is, largely, variations of the old theme,
the gravity law anticipated by Hook and derived from Kepler's laws
by Newton:
 \be
m_i\;{\bf \ddot { \robold}}_i\;=\;m_i\left\{ \sum_{j \neq
i}\,\;G\,m_j\;\frac{{\bf {\erbold}}_{ij}}{r_{ij}^3}
\right\}\;=\;m_i\;\frac{\partial U_i}{\partial {\bf
{\robold}}_i}\;\;\;,\;\;\;\;\;{\bf {\erbold}}_{ij}\,\equiv\,{\bf
{\robold}}_j\,-\,{\bf {\robold}}_i\;\;,
\;\;\;\;i,j=1,...,N\;\;,\;\;\;
 \label{2.1}
  \ee
 $m_i$ and
${\robold}_i$ being the masses and their inertial-frame-related
positions, $\,U_i$ being the overall potential acting on $m_i$:
 \be
  U_i\;\equiv\;\sum_{j
\neq i}\,\;G\,m_j\;\frac{1}{\rho_{ij}}\;\;,
 \label{2.2}
 \ee
the sign convention chosen as in the astronomical, not as in the
physical literature. The equations of motion may be conveniently
reformulated in terms of the relative locations
 \be
 { \erbold }_i\;\equiv\;\erbold_{is}\;\equiv\;{\bf {\robold}}_i\;-\;{\bf {\robold}}_s\;\;\;,
  \label{2.3}
   \ee
${\bf {\robold}}_s\;$ standing for the position of Sun. The
difference between
 \be
  {\bf \ddot {\robold}}_i\;=\;\sum_{j\neq i,
s}\;\;G\;
\frac{m_j\;{\erbold}_{ij}}{r_{ij}^3}\;+\;G\;\frac{m_s\;{\bf
{\erbold}}_{is}}{r_{is}^3}\;
 \label{2.4}
  \ee
and
 \be
  {\bf \ddot {\robold}}_s\;=\;\sum_{j\neq i, s}\;\;G\;
\frac{m_j\;{\bf {\erbold}}_{sj}}{r_{sj}^3}\;+\;G\;\frac{m_i\;{\bf
{\erbold}}_{si}}{r_{si}^3}\;
 \label{2.5}
 \ee
amounts to:
 \be
{\mbox{\boldmath$\ddot {\erbold}$}}_i\;=
 \;
 \sum_{j\neq i,s}\;\;G\;\frac{m_j\;{\bf {\erbold}}_{ij}}{r_{ij}^3}
 \;-\;
 \sum_{j\neq i,s}\;\;G\;\frac{m_j\;{\erbold}_{j}}{r_{j}^3}
 \;-\;
 G\;\frac{\left(m_i\,+\,m_s\right)\;{\erbold}_{i}}{r_{i}^3}
 \;=\;
 \frac{\partial\,\tilde{U}_{i}}{\partial\,{\erbold}_{i}}
 \label{2.6}
 \ee
$\tilde U\;$ being the new potential:
 \be
 \tilde{U}_{i}\;\equiv\;\frac{G\;\left(m_i\;+\;m_s\right)}{r_i}\;+\;R_i\;\;,
 \label{2.7}
 \ee
with the disturbing function
 \be
 R_i\;\equiv\;\sum_{j\neq i}\;\;G\;m_j\;\left\{\frac{1}{r_{ij}^3}\;
 -\;\frac{{\bf {\robold}}_i\,\cdot\,{\bf {\robold}}_j }{\rho_j^3}   \right\}
 \label{2.8}
 \ee
singled out.

Formulae (\ref{2.6}) - (\ref{2.7}) become trivial in the case of
the two-body problem where only $\;m_{\it i}\;$ and $\;m_s\;$ are
present. In this situation the disturbing function vanishes and
the motion is, mathematically, equivalent to rotation about a
nailed-down body of mass $\;m_{\it i}\,+\,m_s\; $ located at some
fixed point $\,O\,$:
 \be
  {\mbox{\boldmath$\ddot {\erbold}$}}\;+\;\frac{\mu}{r^2}\;\frac{{\mbox{\boldmath$\vec
r$}}}{r}\;=\;0
\;\;\;\;,\;\;\;\;\;\;\;\erbold\;\equiv\;\robold_{planet}\;-\;\robold_{s}\;\;\;,\;\;\;\;\;\;\;\mu\;\equiv\;G\;\left(m\;+\;m_s
\right)\;\;\;.
 \label{2.9}
  \ee
In here $\;{\mbox{\boldmath$\vec
r$}}\,\equiv\,{\mbox{\boldmath$\vec r$}}_1\,\equiv\,
{\mbox{\boldmath$\vec r$}}_i\;$, because the subscript $\;i\;$
runs through one value solely: $\;i\,=\,1\;$.

This setting permits exact analytical treatment that leads to the
famous Newtonian result: the orbit is elliptic and has the
gravitating centre in one of its foci. This enables a transition
from Cartesian to Keplerian coordinates. For our further study
this transition will be very important, so we shall recall it in
detail.

At any instant of time, the position $\; {\mbox{\boldmath$r$}} \;$
and velocity $\;\dot {\mbox{\boldmath$r$}}
 \;$ of an orbiting body can be determined by its coordinates
$\;(x,\,y,\,z)\;$ and derivatives $\;(\dot x,\, \dot y,\,\dot
z)\;$ in an inertial frame with origin located in point $\,O\,$
where the  mass $\;m_{\it i}\,+\,m_s\;$ rests. The position of the
orbital ellipse may be fully defined by the longitude of the node,
$\;\Omega\;$; the inclination, $\;\inc\;$; and the argument of
pericentre, $\;\omega\;$ (instead of the latter, one can introduce
the longitude of pericentre,
$\;\tilde{\omega}\,\equiv\,\Omega\,+\,\omega\; $). The shape of
the ellipse is parametrised by its eccentricity, $\;e\;$, and
semimajor axis, $\;a\;$. Position of a point on the ellipse may be
charachterised, for example, by the eccentric anomaly, $\;E\;$. As
well known,
 \be
 E\;-\;e\;\sin E\;=\;n\;t\;-\;B\;\;\;,
 \label{2.10}
 \ee
$B\;$ being a constant of integration, and $\;n\;$ being the mean
motion defined as
 \be
 n\;\equiv\;\mu^{1/2}\;a^{-3/2}\;\;.
 \label{2.11}
 \ee
One can then introduce, following Kepler, the mean anomaly,
$\;M\;$ as
 \be
 M\;\equiv\;E\;-\;e\;\sin E\;\;\;.
 \label{2.12}
 \ee
Let $\,t_o\,$ be the fiducial time. Then, by putting
$\,B\,=\,M_o\,+\,n\,t_o\,$, we can introduce, instead of $\,B\,$,
another integration constant, $\,M_o\,$. Hence, (\ref{2.10}) will
read:
 \be
 M\;=\;M_o\;+\;n\;\left(t\,-\,t_o \right)\;\;,
 \label{2.13}
 \ee
the meaning of $M_o$ being self-evident: it is the value of $M$ at the
reference epoch $\,t_o$. So introduced the mean anomaly provides another
parameterisation of the position of a planet on the ellipse. In the disturbed
case the latter formula naturally becomes
 \be
 M\;=\;M_o\;+\;\int_{t_o}^t n(t')\,dt'\;\;.
 \label{2.131}
 \ee
One more convenient parameter often employed in the literature is
the mean longitude $\,\lambda\,$ defined by
 \be
 \lambda\;\equiv\;\tilde \omega\;+\;{\it
M}\;=\;\Omega\,+\,\omega\;+\;{\it M}\; \;.
 \label{2.14}
 \ee
 Unless the inclination is zero, neither the longitude of the pericentre, $\,\tilde{\omega}\,$,
 nor the mean longitude, $\,\lambda\,$, is a true angle. They are sums of angles in two
 different planes that meet at the node.

Planetary dynamics is based on application of the above, 2-body,
formalism to the N-body case. Naively speaking, since the mutual
disturbances of planets are very weak compared to the solar
gravity, it seems natural to assume that the planets move along
ellipses which are slowly evolving. Still, the weakness of
perturbations is, by itself, a very shaky foundation for the
varying-ellipse method. This so physically-evident circumstance
has a good illustrative power but is of no help when the following
questions arise:

(1) To what degree of rigour can an orbit curve be modelled by a
family of instantaneous ellipses having the Sun in one of their
foci? Can this be performed exactly?

(2) Is this representation of the curve by a family of ellipses unique?

These two questions will not seem anecdotal, if we recall that the
concept of evolving instantaneous ellipses had been introduced
into practice (and that major developments of the
disturbing-function theory had been accomplished) long before
Frenet and Serait developed the theory of curves.\footnote{I am
grateful to William Newman who drew my attention to this
circumstance.} (This historical paradox explains the reason why
words "helicity" and "torsion" are still absent from the astronomy
textbooks.)

Fortunately, Lagrange, who authored the idea of instantaneous
ellipses, fortified it with such powerful tools of calculus, that
in this case they surpassed the theory of curves. Moreover, these
tools in no way relied on the weakness of the disturbances. Hence,
Lagrange's treatment of the problem already contained an
affirmative answer to the first question.

Below we shall demonstrate that the answer to the second question
is negative. Moreover, it turns out that the question calls into
being a rich, though not new, mathematical structure. We shall
show that the Lagrange system of equations for the instantaneous
orbital elements possesses a hidden symmetry not visible with the
naked eye. This symmetry is very similar to the gauge symmetry,
one well known from electrodynamics. A careful analysis shows that
the Lagrange system, as we know it, is written in some specific
gauge: all trajectories constrained to some 9-dimensional
submanifold in the 12-dimensional space constituted by the
Keplerian elements and their time derivatives.

Beside the possible practical relevance to orbit computation, the
said symmetry unveils a fiber bundle structure hidden behind
Lagrange's system of equations for the Keplerian elements. The
symmetry is absent in the 2-body case, but comes into being in the
N-body setting ($N\,\ge \,3$) where each orbiting body follows a
ellipse of varying shape, but the time evolution of the ellipse
contains an inherent ambiguity.

Here follows a crude illustration of this point. Imagine two coplanar ellipses
sharing one focus. Let one ellipse slowly rotate within its plane, about the shared focus.
Let the other ellipse rotate faster, also in its plane, in the
same direction, and about that same shared focus. Suppose a planet is at one of
 the points of these ellipses'
intersection. One observer may state that the planet is rapidly moving along a
slowly rotating ellipse, while another observer may insist that the planet is
slowly describing the fast-moving ellipse. Both descriptions will be equally
legitimate, for there exists an infinite number of ways of dividing the actual
motion of the planet into its motion along some orbit and simultaneous
evolution of the orbit itself. Needless to say, the real, physical trajectory
is unique. However, its description (parameterisation in terms of Kepler's
elements) is not. A map between two different (though physically-equivalent)
sets of orbital elements is a symmetry transformation (a gauge transformation,
in physicists' jargon).

Lagrange never dwelled on that point. However, in his treatment he
passingly introduced a convenient mathematical condition similar
to (\ref{1.5}), which removed the said ambiguity. This condition
and possible alternatives to it will be the topic of the further
sections.

\pagebreak

\section{Keplerian coordinates in the 2-body and N-body problems:
Osculating Elements vs Orbital Elements}

Although not widely recognised, the perturbation equations of
celestial mechanics possess a gauge freedom. It is probable that
this was already noticed by Euler and Lagrange in the middle of
the XVIII century. However, although the existence of this freedom
did not entirely escape attention,\footnote{~In 2002 a veteran of
the Russian astronomy Yurii Batrakov mentioned to the author that
back in his student years his lecturer at the St.Petersburg (then
Leningrad) State University, Mihail Subbotin, emphasised in his
course the possibility of making alternative choices.
Interestingly, though, this fact was not developed in the
published version of that lecture course (Subbotin 1968) where it
is only passingly mentioned that the Lagrange constraint is
imposed merely for convenience.} its consequences have yet to be
fully explored.

Perhaps the easiest way to gain an appreciation of this freedom is
to follow the derivation of the perturbation equations by
application of the variation of parameters (VOP) technique as
invented by Euler and Lagrange and shaped into its final form in
Lagrange (1808, 1809, 1810). Lagrange put the planetary equations
in a closed form in which temporal derivatives of the orbital
elements were expressed in terms of partial derivatives of the
disturbing function with respect to the orbital elements. A
closely related development is presented in the textbook by
Brouwer and Clemence (1961). We shall start in the spirit of
Lagrange but will soon deviate from it in two points. First, in
this subsection we shall not assume that the disturbing force is
conservative and that it depends upon the positions solely, but
shall permit it to depend also upon velocities. Second, neither in
this subsection nor further shall we impose the Lagrange
constraint.

Before addressing the N-particle case, Lagrange referred to the
reduced 2-body problem,
 \ba
 \nonumber
{{\bf {\ddot {\vec {r}}}}}\;+\;\frac{\mu}{r^2}\; \frac{{{ \bf \vec
r}}}{r}\;=\;0\;\;\;,
\;\;\;\;\;\;\;\;\;\;\;\;\;\;\;\;\;\;\;\;\;\;\;\;\;\;\;\;\\
\label{1}\\
 \nonumber \erbold\;\equiv\;\erbold_{planet}\,-\;\erbold_{sun}
 \;\;\;,\;\;\;\;\;\;\;\;\;\;\;\;\;\mu\;\equiv\;G(m_{planet}\,+
 \,m_{sun})\;\;\;.
  \ea
whose generic solution, a Keplerian ellipse or a hyperbola,
 can be expressed, in some fixed Cartesian frame, as
  \ba
 \nonumber x\;=\;f_1\left(C_1, ... , C_6, \,t
 \right)\;\;\;,\;\;\;\;\;\;\;\;\;\;\dot x\;=\;g_1\left(C_1, ... , C_6,
 \,t \right)\;\;\;\;,\;\\ y\;=\;f_2\left(C_1, ... , C_6, \,t
 \right)\;\;\;,\;\;\;\;\;\;\;\;\;\;\dot y\;= \;g_2\left(C_1, ... ,
 C_6, \,t \right)\;\;\;\;,\;
 \label{2}\\
 \nonumber z\;=\;f_3\left(C_1, ... , C_6, \,t
 \right)\;\;\;,\;\;\;\;\;\;\;\;\;\;\dot z\;= \;g_3\left(C_1, ... ,
 C_6, \,t \right)\;\;\;\;,\;\,
  \ea
   or, shortly:
    \be
    \erbold \;=\; {\bf
 \vec f} \left(C_1, ... , C_6, \,t \right)\;\;\;,\;\;\;\;\;\;
 \;\;\;\;{\bf \dot{\erbold}} \;=\;{\bf \vec g}\left(C_1, ... , C_6,
 \,t \right) \;\;\;\;\;\;,\;\;\;\;\;\;
 \label{3}
 \ee
the functions $g_i$ being, by definition, partial derivatives of
$f_i$ with respect to the last argument:
 \be
 {{\bf{\vec{g}}}}\;\equiv\;\left(\frac{\partial {\bf {\vec f}}}{
 \partial
 t}\right)_{C=const} .
 \label{4}
 \ee
Naturally, the general solution contains six adjustable constants,
$C_i$, since the problem (\ref{1}) is constituted by three, second
order, differential equations.  To find the explicit form of the
dependence (\ref{3}), one can employ an auxiliary set of Cartesian
coordinates $\;\bf{\vec q}\;$, with an origin at the gravitating
centre, and with the first two axes located in the plane of orbit.
In terms of the true anomaly $\;\it f\;$, these coordinates will
read:
 \ba
q_1\;\equiv\;r\;\cos f\;\;,\;\;\;q_2\;\equiv\;r\;\sin
f\;\;,\;\;\;q_3\;=\;0\;\;.
 \label{5}
 \ea
In the two-body case, their time derivatives can be easily
computed and expressed through the major semiaxis, $\;a\;$, the
eccentricity, $\;e\;$, and the true anomaly, $\;\it f\;$:
 \ba
\dot{q}_1\;=\;-\,\frac{n\;a\;\sin
f}{\sqrt{1\,-\,e^2}}\;\;,\;\;\;\dot{q}_2\;=\;\frac{n\;a\;(e\,+\,\cos
f)}{\sqrt{1\,-\,e^2}}\;\;,\;\;\;\dot{q}_3\;=\;0\;\;.
 \label{6}
 \ea
 (true anomaly $\;\it f\;$ itself being a function of $\;a\;$,
$\;e\;$, and of the mean anomaly
$\;M\,\equiv\,M_o\,+\,\int_{t_o}^t n\;dt\;$,
$\;\;n\,\equiv\,\mu^{1/2}a^{-3/2}\;$). In the two-body setting,
the inertial-frame-related position and velocity will appear as:
 \ba
 \nonumber
{{\erbold}} \;=\;{\bf R}(\Omega,\,\inc,\,\omega)\;\,{\bf {\vec
q}}(a,\,e,\,M_o\,,\,t)\;\;\;,\\
 \label{7} \\
 \nonumber
 {\bf {\dot{\erbold}}}\;=\;{\bf R}(\Omega,\,\inc,\,\omega)\;\,{\bf
{\dot{\vec q}}}(a,\,e,\,M_o\,,\,t)\;\;\;\;,
 \ea
 ${\bf R}(\Omega,\,\inc,\,\omega)\;$ being the matrix of rotation
from the orbital-plane-related axes $\;\bf q\;$ to the fiducial
frame $\;(x,\,y,\,z)\;$ in which the vector $\;\bf \vec r\;$ is
defined. The rotation is parametrised by the three Euler angles:
inclination, $\;\inc\;$; the longitude of the node, $\;\Omega\;$;
and the argument of the pericentre, $\;\omega\,$.

This is one possible form of the general solution (\ref{3}). It
has been obtained under the convention that a particular ellipse
is parametrised by the Lagrange set of orbital elements,
$\,C_i\,\equiv\,(e,\,a,\,M_o,\,\omega,\,\Omega,\,i)\,$. A
different functional form of the same solution is achieved in
terms of the Delaunay set,
$\,D_i\,\equiv\,(L,\,G,\,H,\,\omega,\,\Omega,\,M_o)\,$. Still
another possibility is to express the general solution through the
initial conditions: then the constants
$\,(x_o,\,y_o,\,z_o,\,\dot{x}_o,\,\dot{y}_o,\,\dot{z}_o)\,$ are
the six parameters defining a particular orbit. The latter option
is natural when the integration is carried out in Cartesian
components, but is impractical otherwise.

At this point it is irrelevant which particular set of the
adjustable parameters is employed. Hence we shall leave, for a
while, the solution in its most general form (\ref{2} - \ref{4})
and shall, following Lagrange (1808, 1809, 1810), employ it as an
ansatz for a solution of the N-particle problem where the
disturbing force acting at a particle is denoted by
$\;\Delta{\bf{\vec F}}\,$:\footnote{Our treatment covers
disturbing forces $\Delta{\bf\vec F}$ that are arbitrary
vector-valued functions of position and velocity.}
 \be {\bf \ddot{\vec
r}}\;+\;\frac{\mu}{r^2}\;\frac{{\bf\vec r }}{r}\;=\;{\Delta
\bf{\vec F}}\;\;\;,
 \label{8}
 \ee
the "constants" now being time dependent:
 \be
{\bf \vec r }\;=\;{\bf \vec f} \left(C_1(t), ... , C_6(t), \,t
\right)\,\;\;\;,
 \label{9}
 \ee
and the functional form of $\;\bf{\vec{f}}\;$ remaining the same
as in (\ref{3}). Now the velocity
 \be
 \frac{d \erbold}{dt}\;=\;\frac{ \partial {\bf \vec
f}}{\partial t}\;+\; \sum_i \;\frac{\partial {\bf \vec
f}}{\partial C_i}\;\frac{d C_i}{d t}\;= \;{\bf \vec g}\;+\; \sum_i
\;\frac{\partial {\bf \vec f}}{\partial C_i}\;\frac{d C_i}{d
t}\;\;\;\;,
 \label{10}
 \ee
will contain a new input, $\;\sum ({\partial{\bf \vec
f}}/{\partial C_i})({dC_i}/{dt})\;$, while the first term, $\;\bf
{\vec{g}}\;$, will retain the same functional form as it used to
have before.

Substitution of $\; {\bf \vec f} \left(C_1(t), ... , C_6(t), \,t \right)\,$
into the perturbed equation of motion (\ref{8}) gives birth to three
independent scalar differential equations of the second order. These three
equations contain one independent parameter, time, and six time-dependent
variables $\;C_i( t)\;$ whose evolution is to be determined. Evidently, this
cannot be done in a single way because the number of variables exceeds, by
three, the number of equations. This means that, though the {\it{~physical~}}
orbit (comprised by the locus of points in the Cartesian frame and by the
values of velocity in each of these points) is unique, its parameterisation in
terms of the orbital elements is ambiguous. Lagrange, in his treatment,
noticed that the system was underdefined, and decided to amend it with exactly
three independent conditions which would make it solvable.

Before moving on with the algebra, let us look into the
mathematical nature of this ambiguity. A fixed Keplerian ellipse
(\ref{3}), which is the solution to the 2-body problem (\ref{1}),
gives birth to a time-dependent one-to-one (within one revolution
period) mapping
 \be
  \; \left(\; C_{1} \, , \;...\; ,
\;C_{6}\;\right) \;\longleftrightarrow\; (\; x(t) \, , \; y(t)\, ,
\; z(t) \, , \; \dot{x}(t) \, , \; \dot{y}(t) \, , \; \dot{z}(t)
\; ) \;\;\;.
 \label{3.5}
 \ee
 In the N-body case, the new ansatz (\ref{9})
  is incompatible with
(\ref{3}). This happens because now the time derivatives of
coordinates $\;C_i\;$ come into play in (\ref{10}). Hence, instead
of (\ref{3.5}), one gets a time-dependent mapping between a
12-dimensional and a 6-dimensional spaces:
 \be
  \;\left(\;C_1(t)\,,\; ...\; ,\; C_6(t)\,, \; \dot{C}_{1}(t)\,,\;
 ...\; ,\; \dot{C}_6(t)\;\right) \;\rightarrow\;
 (\;x(t)\,,\;y(t)\,,\;z(t)\,,\;\dot{x}(t)\,,\;\dot{y}(t)\,,\;\dot{z}(t)\;)
 \,\;.
 \label{3.8}
 \ee
This brings up two new issues. One is the multiple scales: while the physical
motion along an instantaneous ellipse parametrised by $\,C_i\, $ is associated
with the "fast time scale", the evolution of the osculating elements
$\,C_i(t)\,$ represents the "slow time scale". The quotation marks are
necessary here, because in reality these time scales are inseparably
connected, so that the "slow time scale" is not always slower than the "fast
time scale". What is important here is that, in general, ansatz (\ref{9})
gives birth to two separate time scales. \footnote{In practice, the mean
longitude $\,\lambda\,=\,\lambda_o\,+\,\int_{ t_o}^{t}n(t)\,dt\;$ is often
used instead of its fiducial-epoch value $\, \lambda_o\,$. Similarly, those
authors who prefer the mean anomaly to the mean longitude, often use
$\,M\,=\,M_o \,+\,\int_{t_o}^{t}n(t)\,dt\;$ instead of $\, M_o\,$. While
$\,M_o\,$ and $\,\lambda_o\,$ are orbital elements, the quantities $\,M\,$ and
$\,\lambda\,$ are not. Still, the time-dependent change of variables from
$\,\lambda_o\,$ to $\,\lambda\,$ (or from $\,M_o\,$ to $\,M \,$) is perfectly
legitimate. Being manifestly time-dependent, this change of variables
intertwines two different time scales: for example, $\,M\,$ carries a ``fast''
time dependence through the upper limit of the integral in $\,M\,=\,
M_o\,+\,\int_{t_o}^{t}n(t)\,dt\;$, and it also carries a ``slow''
time-dependence due to the adiabatic evolution of the osculating element
$\,M_o\,$. The same concerns $\,\lambda\,$.} The second important issue is
that mapping (\ref{3.8}) cannot be one-to-one. As already mentioned, the three
equations of motion (\ref{8}) are insufficient for determination of six
functions $\;C_1,\,...\,C_6\;$ and, therefore, one has a freedom to impose, by
hand, three extra constraints upon these functions and their
derivatives\footnote{A more accurate mathematical discussion of this freedom
should be as follows. The dynamics, in the form of first-order differential
equations for the orbital coordinates $\;C_i(t)\;$ and their derivatives
$\;H_i(t )\,\equiv\,{\bf{\dot{\rm{C}}}}_i(t)\;$, will include six evident
first-order identities for these twelve functions:
$\;H_i(t)\;=\;dC_i(t)/dt\;$. Three more differential equations will be
obtained by plugging $\,\erbold = \efbold (C_1,...,C_6,t)\,$ into (\ref{8}).
These equations will be of the second order in $\;C_i(t)\;$. However, in terms
of both $\;C_i(t)\;$ and $\;H_i(t)\;$ these equations will be of the first
order only. Altogether, we have nine first-order equations for twelve
functions $\;C_i(t)\;$ and $\;H_i(t)\;$. Hence, the problem is underdefined
and permits three extra conditions to be imposed by hand. The arbitrariness of
these conditions reveals the ambiguity of the representation of an orbit by
instantaneous Keplerian ellipses. Mappings between different representations
reveal an internal symmetry (and a symmetry group) underlying this
formalism.}.

Though Lagrange did notice that the system was underdefined, he
never elaborated on the underlying symmetry. He simply imposed
three convenient extra conditions
 \be
  \sum_i \;\frac{\partial
{\bf{\vec{f}}}}{\partial C_i}\;\frac{d C_i}{d t}\;=\;0 \;\;\;,
 \label{3.9}
  \label{12}
 \label{3.15}
  \ee
 and went on, to derive (in this particular gauge,
which is often called "Lagrange constraint") his celebrated system
of equations for orbital elements. Now we can only speculate on
why Lagrange did not bother to explore this ambiguity and its
consequences. One possible explanation is that he did not have the
concept of continuous groups and symmetries in his arsenal (though
it is very probable that he knew the concept of discrete
group\footnote{In his paper on solution, in radicals, of equations
of degrees up to four, {\it{R{\'{e}}flexions sur la
r{\'{e}}solution alg{\'{e}}brique des {\'{e}}quations}}, dated by
1770, Lagrange performed permutations of roots. Even though he did
not consider compositions of permutations, his technique reveals
that, most likely, he was aware of, at least, the basic idea of
discrete groups and symmetries.}). Another possibility is that
Lagrange did not expect that exploration of this ambiguity would
reveal any promising tools for astronomical calculations.

Lagrange's choice of the supplementary constraints was motivated
by both physical considerations and the desire to simplify
calculations. Since, physically, the time-dependent set
$\,\left(C_1(t), ... , C_6(t)\right)\,$ can be interpreted as an
instantaneous ellipse, in a bound-orbit case, or an instantaneous
hyperbola, in a fly-by situation, Lagrange decided to make the
instantaneous orbital elements $\;C_i\;$ osculating, i.e., he
postulated that the instantaneous ellipse (or hyperbola) must
always be tangential to the physical trajectory. This means that
the physical trajectory defined by $\,\left(C_1(t), ... ,
C_6(t)\right)\,$ must, at each instant of time, coincide with the
unperturbed (two-body) orbit that the body would follow if
perturbations were to cease instantaneously. This can be achieved
only in case the velocities depend upon the elements, in the
N-body problem, in the same manner as they did in the 2-body case.
This, in turn, can be true only if one asserts that the extra
condition (\ref{12}) is fulfilled. That condition, also called
Lagrange constraint, consists of three scalar equations which,
together with the three equations of motion (\ref{8}), constitute
a well-defined system of six equations with six variables
$\,\left(C_1(t), ... , C_6(t)\right)\,$.

For the reasons explained above, this constraint, though well
motivated, remains, from the mathematical viewpoint, completely
arbitrary: it considerably simplifies the calculations but does
not influence the shape of the physical trajectory and the rate of
motion along that curve. One could as well choose some different
supplementary condition
 \be
\sum_i \;\frac{\partial {\bf \vec f}}{\partial C_i}\;\frac{d
C_i}{d t}\;=\; {\bf {\vec \Phi}}(C_{1,...,6}\,,\,t)\;\;\;,\;
 \label{13}
  \ee $\Phibold\,$ now being an arbitrary function of time and
parameters $\;C_i\;$.\footnote{In principle, one may endow $\,\Phibold\,$ also
with dependence upon the parameters' time derivatives of all orders. This
would yield higher-than-first-order time derivatives of the $\;C_i\;$ in
subsequent developments requiring additional initial conditions, beyond those
on ${\bf\vec r}$ and ${\bf\dot{\vec r}}$, to be specified in order to close
the system. We avoid this unnecessary complication by restricting $\Phibold\,$
to be a function of time and the $\;C_i\;$.} Substitution of (\ref{12}) by
(\ref{13}) would leave the physical motion unchanged but would alter the
subsequent mathematics and, most importantly, would eventually yield different
solutions for the orbital elements.  Such invariance of a physical theory
under a change of parameterisation is an example of gauge symmetry (Efroimsky
2002).  This freedom is a particular case of a more general type of invariance
that emerges in applications of the VOP method to a vast variety of ODE
problems.

The importance of this gauge freedom is determined by two circumstances which
parallel similar circumstances in electrodynamics. One consequence of the
gauge invariance is its non-conservation in the course of orbit computation.
This, purely numerical, phenomenon may be called gauge drift. Another relevant
issue is that a clever choice of gauge often simplifies the solution of the
equations of motion. In application to the theory of orbits, this means that a
deliberate choice of non-osculating orbital elements (i.e., of a set $\;C_i\;$
obeying some condition (\ref{13}) different from (\ref{12})) can sometimes
simplify the equations for these elements' evolution. As explained in
Efroimsky \& Goldreich (2003b), some earlier developments by Goldreich (1965),
Brumberg et al (1971) can be interpreted as calculations in non-Lagrange
gauges.

The standard derivation of both Delaunay and Lagrange systems of planetary
equations by the VOP method rests on the assumption that the Lagrange
constraint (\ref{12}) is fulfilled. Both systems get altered under a different
gauge choice. Derivation of the gauge-invariant (i.e., valid in an arbitrary
gauge (\ref{13})) Lagrange system can be found in Efroimsky (2002). Derivation
of the gauge-invariant Delaunay system is given in Efroimsky (2002). In
Efroimsky \& Goldreich (2003b) both systems are amended with extra terms which
emerge when the disturbance happens to depend not only upon the positions but
also upon velocities. This situation is encountered when one is working in a
non-inertial coordinate frame or when relativistic corrections to the
Newtonian force are accounted for.

Without going into excessive details, that may be found in
(Efroimsky 2002), we give the essence of that derivation so that
we can refer to it in one of the subsequent sections.

>From (\ref{10}), it follows that the formula for the acceleration
reads:
 \be
 \frac{d^2 \erbold
}{dt^2}\;= \;\frac{\partial \bf \vec g}{\partial
t}\;+\;\sum_i\;\frac{\partial{\bf \vec g}}{\partial C_i}\;\frac{d
C_i}{d t}\;+\;\frac{d{\bf \vec \Phi}}{dt}\;=\; \frac{\partial^2
{\bf{\vec f}}}{\partial^2 t}\;+\;\sum_i\;\frac{\partial{\bf \vec
g}}{\partial C_i }\;\frac{d C_i}{d t}\;+\;\frac{d{\bf \vec
\Phi}}{dt}\;\;,
 \label{210}
 \ee
Together with the equation of motion (\ref{8}), it leads to:
 \be
\frac{\partial^2 {\bf {\vec f}} }{\partial
t^2}\;+\;\frac{\mu}{r^2}\;\frac{\bf{\vec f}}{r} \;+\;\sum_i\;
\frac{\partial {\bf \vec g}}{\partial C_i}\;\frac{d C_i}{d t}\;+\;
\frac{d{\bf \vec \Phi}}{dt}\;=\;\Delta{\bf{\vec
F}}\;\;\;\;,\;\;\;\;\;\;r\;\equiv\;|\erbold|\;=\;|{\bf {\vec
f}}|\;\;\;.
 \label{211}
 \ee
As $\;\bf {\vec f}\;$ is, by definition, a Keplerian solution to
the two-body problem (and, therefore, obeys the unperturbed
equation (\ref{1})$\,$), the above formula becomes:
 \be
 \sum_i\; \frac{\partial {\bf
\vec g}}{\partial C_i}\;\frac{d C_i}{d t}\;=\;{\Delta{\bf{\vec
F}}}\;-\; \frac{d{\bf{\vec{\Phi}}}}{dt}\;\;\;\;.
 \label{212}
 \ee
This is the equation of disturbed motion, written in terms of the
orbital elements. Together with constraint (\ref{13}) it
constitutes a well-defined system of equations that can be solved
with respect to $\;dC_i/dt\;$ for all $i$'s. The easiest way of
doing this is to employ the elegant technique offered by Lagrange:
to multiply the equation of motion by $\;\partial{\bf{\vec
f}}/\partial{C_n}\;$ and to multiply the constraint by
$\;-\,\partial{\bf{\vec g} }/\partial{C_n}\;$. These operations
yield the following equalities
 \be
 \frac{\partial \bf{\vec f}}{\partial
C_n} \,\left( \sum_j \, \frac{\partial \bf \vec g}{\partial
C_j}\,\frac{dC_j}{dt} \right)\;=\;\frac{\partial \bf{\vec
f}}{\partial C_n}\;{\;\Delta{\bf{\vec F}}}\;-\;\frac{\partial
\bf{\vec f}}{\partial C_n}\;\frac{d \bf \vec \Phi}{dt} \;
 \label{214}
 \ee
and
 \be
  -\;\frac{\partial \bf \vec g}{\partial C_n}\,
  \left( \sum_j\,\frac{\partial \bf \vec f}{\partial C_j}\,
  \frac{dC_j}{dt} \right)\;=\;-\;\frac{\partial \bf \vec
g}{\partial C_n} \;{\bf \vec \Phi} \;\;\;,
 \label{215}
 \ee
summation whereof results in:
 \be
\sum_j\;[C_n\;C_j]\;\frac{dC_j}{dt}\;=\;
 \frac{\partial \bf {\vec f}}{\partial C_n}\;
  {\Delta \bf{\vec F}}\;-\; \frac{\partial{\bf
{\vec f}}}{\partial C_n} \;\frac{d \bf {\vec \Phi}}{dt}\;-\;
\frac{\partial  \bf {\vec g}}{\partial C_n} \;{\bf {\vec \Phi}}
\;\;\;\;,
 \label{217}
 \ee
where the symbol $\;[C_n\;C_j]\;$ denotes the unperturbed (i.e.,
defined as in the two-body case) Lagrange brackets:
 \be
[C_n\;C_j]\;\equiv\;\frac{\partial {{\bf{\vec f}}}}{\partial
C_n}\, \frac{\partial {\bf {{\vec g}}}}{\partial
C_j}\,-\,\frac{\partial {{\bf{\vec f}}}}{\partial C_j}\,
\frac{\partial {\bf {{\vec g}}}}{\partial C_n}\;\;\;\;.
 \label{218}
 \ee
Above we specified that $\;\Phibold\;$ is a function of time and
the parameters $\;C_i\;$, but not of their derivatives. Under this
restriction, (\ref{217}) may be conveniently rewritten as
 \be
\sum_j\;\left(\,[C_n\;C_j]\;+\;\frac{\partial \bf\vec f}{\partial
C_n}\;\frac{\partial \bf {\vec \Phi}}{\partial C_j}\;
\,\right)\,\frac{dC_j}{dt}\;=\; \frac{\partial \bf {\vec
f}}{\partial C_n}\; {\Delta \bf{\vec F}}\;-\; \frac{\partial{\bf
{\vec f}}}{\partial C_n} \;\frac{\partial \bf {\vec
\Phi}}{\partial t}\;-\; \frac{\partial \bf {\vec g}}{\partial C_n}
\;{\bf {\vec \Phi}} \;\;\;\;.
 \label{general_F}
 \ee
This expresses the most generic form of the gauge-invariant
perturbation equations of celestial mechanics, that follows from
the VOP method.\footnote{~I am grateful to Peter Goldreich who
recommended me to present the equations in this concise and so
general form.}

We can obtain an immediate solution for the individual $dC_i/dt$
in the Lagrange gauge by exploiting the well known expression for
the inverse Lagrange-bracket, or Poisson-bracket, matrix that is
offered in the literature for the two-body problem. The presence
of the term proportional to $\partial{\vec \Phi}/ \partial C_j$ on
the lhs of (\ref{general_F}) complicates the solution for the
individual $dC_i/dt$ in an arbitrary gauge, but only to the extent
of requiring the resolution of a set of six, simulataneous,
linear, algebraic equations.

All different choices of three (compatible and sufficient) gauge conditions
expressed by the vector $\;\Phibold\;$ will lead to physically equivalent
results. This equivalence means the following. Suppose we solve the equations
of motion for $\;C_{1,...,6}\;$, with some gauge condition $\,{\bf \vec
\Phi}\,$ enforced. This will give us the solution, $\;C_{1,...,6}(t) \;$. If,
though, we choose to integrate the equations of motion with another gauge
$\,\tilde{\Phibold}\,$ enforced, then we shall arrive at a solution
$\;{\tilde{C}}_{i}(t)\;$ of a different functional form. Stated alternatively,
in the first case the integration in the 12-dimensional space
$\;(\;C_{1,...,6}\,,\;H_{1,...,6}\;)\;$ will be restricted to one
9-dimensional time-dependent submanifold, whereas in the second case it will
be restricted to another submanifold. Despite this, both solutions,
$\;C_{i}(t)\;$ and $\;\tilde{C}_{i}(t) \;$, will give, when substituted back
in (\ref{9}), the same orbit $\;(x(t), \,y(t),\,z(t))\;$ with the same
velocities $\;(\dot x(t),\, \dot y(t),\,\dot z(t))\;$. This is a
fiber-bundle-type structure, and it gives birth to a 1-to-1 map of
$\;C_{i}(t)\;$ onto $\;\tilde{C}_{i}(t)\;$. This map is merely a
reparameterisation. In physicists' parlance it will be called gauge
transformation. All such reparameterisations constitute a group of symmetry,
which would be called, by a physicist, gauge group. The real orbit is
invariant under the reparameterisations which are permitted by the ambiguity
of gauge-condition choice. This physical invariance implements itself,
technically, as form-invariance of the expression (\ref{9}) under the afore
mentioned map. This is analogous to Maxwell's electrodynamics: the components
$\;x,\;y,$ and $\,z\,$ of vector $\,\erbold\,$, and their time derivatives,
play the role of the physical fields $\,\bf \vec E\,$ and $\,\bf \vec B\,$,
while the Keplerian coordinates $\,C_1, ... , C_6\,$ play the role of the
4-potential $\,A^{\mu}\,$. This analogy can go even further.\footnote{Suppose
one is solving a problem of electromagnetic wave proliferation, in terms of
the 4-potential $\,A^{\mu}\,$ in some fixed gauge. An analytic calculation
will render the solution in that same gauge, while a numerical computation
will furnish the solution in a slightly different gauge. This will happen
because of numerical errors' accumulation. In other words, numerical
integration will slightly deviate from the chosen submanifold. This effect may
become especially noticeable in long-term orbit computations. Another relevant
topic emerging in this context is comparison of two different solutions of the
N-body problem: just as in the field theory, in order to compare solutions, it
is necessary to make sure if they are written down in the same gauge.
Otherwise, the difference between them may be, to some extent, not of a
physical but merely of a gauge nature.}


\section{The hidden symmetry of the Lagrange system}

If we impose, following Lagrange, the gauge condition
(\ref{3.15}), then the equation of motion (\ref{general_F}) will
simplify:
 \be
\sum_j\;\,[C_n\;C_j]\;\frac{dC_j}{dt}\;=\; \frac{\partial \bf
{\vec f}}{\partial C_n}\; {\Delta \bf{\vec F}}\;\;\;\;\;.
 \label{4.1}
 \ee
In assumption of the disturbing force being dependent only upon
positions and being conservative,\footnote{~I am grateful to Peter
Goldreich who drew my attention to the fact that this assertion is
right in inertial axes solely. Consider an observer who associates
himself with a non-inertial system of reference (say, with a frame
fixed on a precessing planet) and defines the orbital elements in
this system. He will then have to take into account the
non-inertial contribution to the disturbing function. This
contribution, denoted in (Goldreich 1965) by $R_I$, will depend
not only upon $\;{\bf {\vec{r}}}\;$ but also upon $\;{\bf{\dot
{\vec{r}}}}\;$. In this situation, the terms $\;\partial
R/\partial C_r\;$ in the Lagrange equations should be substituted
by $\;\partial R/\partial C_r\;-\;(\partial {\bf{\dot{\vec
r}}}/\partial C_r)(\partial R/\partial {\bf{\dot {\vec{r}}}})$.
The gauge approach to the problem of a satellite orbiting a
wobbling planet will be addressed in one of our subsequent
articles (Efroimsky \& Goldreich 2003b). In regard to the
velocity-dependence of the disturbing function, we would also
mention that a similar situation emerges in the relativistic
mechanics, because the relativistic correction to the Newton law
of gravity bears dependence upon the velocity.} we may substitute
in the above equation the disturbing force by the gradient of a
(position-dependent) disturbing potential:
 \be
 {\Delta \bf{\vec F}}\;=\;\frac{\partial R(\erbold)}{\partial
 \erbold}\;\;\;,
 \label{4.110}
 \ee
which will result in:
 \be
\sum_j\;\,[C_n\;C_j]\;\frac{dC_j}{dt}\;=\; \frac{\partial  {
R}}{\partial C_n}\;\;\;\;\;\;.
 \label{4.111}
 \ee
Expressions for the Lagrange brackets are known (Brouwer \&
Clemence 1961, p.~284), and their insertion into (\ref{4.111})
equation will entail the well-known Lagrange system of planetary
equation:
 \be
  \frac{da}{dt}\;=\;\frac{2}{n\,a}\;\;\frac{\partial R}{\partial
M_o}
 \label{4.6}
 \ee
 \be
\frac{de}{dt}\;=\;\frac{1-e^2}{n\,a^2\,e}\;\;\frac{\partial
R}{\partial M_o}\; \;-\;\;\frac{(1\,-\,e^2)^{1/2}}{n\,a^2\,e}
\;\;\frac{\partial R}{\partial a} \label{4.7} \ee \be
\frac{d\omega}{dt}\;=\;\frac{\;-\;\cos \inc
}{n\,a^2\,(1\,-\,e^2)^{1/2}\, \sin \inc }\;\;\frac{\partial
R}{\partial \inc }\;\;+\;\;
\;\frac{(1-e^2)^{1/2}}{n\,a^2\,e}\;\frac{\partial R}{\partial e}\;
 \label{4.8}
 \ee
 \be \frac{d \inc }{dt}\;=\;\frac{\cos
\inc}{n\,a^2\,(1\,-\,e^2)^{1/2}\, \sin \inc}\;\;\frac{\partial
R}{\partial \omega}\;\;-\;
\;\frac{1}{n\,a^2\,(1\,-\,e^2)^{1/2}\,\sin \inc
}\;\;\frac{\partial R}{\partial \Omega}\;
 \label{4.9}
 \ee
 \be
\frac{d\Omega}{dt}\;=\;\frac{1}{n\,a^2\,(1\,-\,e^2)^{1/2}\,\sin
\inc }\;\; \frac{\partial R}{\partial \inc }\; \label{4.10}
 \ee
 \be
\frac{dM_o}{dt}\;=\;\;-\;\frac{1\,-\,e^2}{n\,a^2\,e}\;\;\frac{\partial
R}{
\partial e }\;\;-\;\;\frac{2}{n\,a}\;\frac{\partial R}{\partial a }\;
 \label{4.11}
 \ee
If analytical integration of this system were possible, it would
render a correct orbit, in the fixed gauge (\ref{3.15}). A
numerical integrator, however, may cause drift from the chosen
submanifold (\ref{3.15}). Even if the drift is not steady, some
deviation from the submanifold is unavoidable. One may wish $\;\bf
\Phi\;$ to be as close to zero as possible, but in reality $\;\bf
\Phi\;$ will be some unknown function whose proximity to zero will
be determined by the processor's error and by the number of
integration steps. Even if we begin with (\ref{3.15}) fulfilled
exactly, the very first steps will give us such values of
$\;C_i\;$ that, being substituted into the lhs of (\ref{3.15}),
they will give some new value of $\;\bf \Phi\;$ slightly different
from zero. Thus, the Lagrange gauge will no longer be observed.

In case we relax the Lagrange constraint and accept an arbitrary
gauge (\ref{13}) then, under the simplifying assertion
(\ref{4.110}), equation (\ref{general_F}) will become
 \be
\sum_j\;\left(\,[C_n\;C_j]\;+\;\frac{\partial \bf\vec f}{\partial
C_n}\;\frac{\partial \bf {\vec \Phi}}{\partial C_j}\;
\,\right)\,\frac{dC_j}{dt}\;=\; \frac{\partial {R}}{\partial
C_n}\;\;-\; \frac{\partial{\bf {\vec f}}}{\partial C_n}
\;\frac{\partial \bf {\vec \Phi}}{\partial t}\;-\; \frac{\partial
\bf {\vec g}}{\partial C_n} \;{\bf {\vec \Phi}} \;\;\;\;.
 \label{general_F}
 \ee
The Lagrange brackets depend exclusively on the functional form of
$x,\,y,\,z\,=\,f_{1,2,3}\left(C_{1,...,6}\,,\;t\right)\;$ and
$\;g_{1,2,3} \equiv{\partial f_{1,2,3} }/{\partial t}\,$, and are
independent from the gauge and from the time evolution of $C_i$.
Hence the gauge-invariant generalisation of the Lagrange system
will emerge:
 \be
\frac{da}{dt}\;=\;\frac{2}{n\,a}\;\;\left[\frac{\partial
R}{\partial M_o} \;-\;{\bf{\vec  \Phi}}\,\frac{\partial \bf {\vec
g}}{\partial M_o} \;-\;\frac{\partial  \bf {\vec f}}{\partial M_o}
\;\frac{d \bf {\vec \Phi}}{dt}
\right]\;\;\;,\;\;\;\;\;\;\;\;\;\;\;\;\;\;\;\;\;\;\;\;\;\;\;\;\;\;\;\;\;\;\;\;\;\;\;\;\;\;\;\;\;\;\;\;\;\;\;\;\;\;\;
 \label{4.15}
 \ee
 \ba
\nonumber
\frac{de}{dt}\,=\,\frac{1-e^2}{n\,a^2\,e}\;\left[\frac{\partial
R}{\partial M_o} \,-\;{\bf {\vec \Phi}}\,\frac{\partial \bf {\vec
g}}{\partial M_o} \;-\;\frac{\partial  \bf {\vec f}}{\partial M_o}
\;\frac{d \bf {\vec \Phi}}{dt}
\right]\;-\;\;\;\;\;\;\;\;\;\;\;\;\;\;\;\;\;\;\;\;\;\;\;\;\;\;\;\;\;\;\;\;\;\;\;\;\;\;\;\;\;\;\;\;\\
\nonumber\\
\frac{(1\,-\,e^2)^{1/2}}{n\,a^2\,e} \;\left[\frac{\partial
R}{\partial a}\;-\;{\bf {\vec \Phi}}\,\frac{\partial \bf {\vec
g}}{\partial a}\;-\;\frac{\partial  \bf {\vec f}}{\partial a}
\;\frac{d \bf {\vec \Phi}}{dt} \right]\;\;,\;
 \label{4.16}
 \ea
 \ba
\nonumber \frac{d\omega}{dt}\;=\;\frac{\;-\;\cos \inc
}{n\,a^2\,(1\,-\,e^2)^{1/2}\, \sin \inc }\;\;\left[\frac{\partial
R}{\partial \inc } \;-\;{\bf {\vec \Phi}}\,\frac{\partial \bf
{\vec g}}{\partial \inc } \;-\;\frac{\partial  \bf {\vec
f}}{\partial \inc } \;\frac{d \bf {\vec \Phi}}{dt}
\right]\;+\;\;\;\;\;\;\;\;\;\;\;\;\;\;\;\;\;\;\;\;\;\;\;\;\;\;\;\;\;\;\; \\
\nonumber\\
\frac{(1-e^2)^{1/2}}{n\,a^2\,e}\;\left[\frac{\partial R}{\partial
e} \;-\;{\bf {\vec \Phi}}\,\frac{\partial \bf {\vec g}}{\partial
e} \;-\;\frac{\partial  \bf {\vec f}}{\partial e} \;\frac{d \bf
{\vec \Phi}}{dt}
\right]\;\;\;,\;\;\;\;\;\;\;\;\;\;\;\;\;\;\;\;\;\;\;\;
 \label{4.17}
 \ea
 \ba
\nonumber \frac{d \inc }{dt}\;=\;\frac{\cos
\inc}{n\,a^2\,(1\,-\,e^2)^{1/2}\, \sin
\inc}\;\;\left[\frac{\partial R}{\partial \omega} \;-\;{\bf {\vec
\Phi}}\,\frac{\partial \bf {\vec g}}{\partial \omega }
\;-\;\frac{\partial  \bf {\vec f}}{\partial \omega } \;\frac{d \bf
{\vec \Phi}}{dt}
\right]\;-\;\;\;\;\;\;\;\;\;\;\;\;\;\;\;\;\;\;\;\;\;\;\;\;\;\;\\
\nonumber\\
\;\frac{1}{n\,a^2\,(1\,-\,e^2)^{1/2}\,\sin \inc
}\;\;\left[\frac{\partial R}{\partial \Omega} \;-\;{\bf {\vec
\Phi}}\,\frac{\partial \bf {\vec g}}{\partial \Omega }
\;-\;\frac{\partial  \bf {\vec f}}{\partial \Omega} \;\frac{d \bf
{\vec \Phi}}{dt} \right]\;\;\;,\;\;
 \label{4.18}
 \ea
 \be
\frac{d\Omega}{dt}\;=\;\frac{1}{n\,a^2\,(1\,-\,e^2)^{1/2}\,\sin
\inc }\;\; \left[\frac{\partial R}{\partial \inc } \;-\;{\bf {\vec
\Phi}}\,\frac{\partial \bf {\vec g}}{\partial \inc }
\;-\;\frac{\partial \bf {\vec f}}{\partial \inc } \;\frac{d \bf
{\vec \Phi}}{dt}
\right]\;\;\;,\;\;\;\;\;\;\;\;\;\;\;\;\;\;\;\;\;\;\;\;\;\;\;\;\;\;\;
 \label{4.19}
 \ee
 \ba
\frac{dM_o}{dt}\,=\,\;-\,\frac{1\,-\,e^2}{n\,a^2\,e}\,\;\left[
\frac{\partial R}{\partial e } \,-\,{\bf {\vec
\Phi}}\,\frac{\partial \bf {\vec g}}{\partial e}
\,-\,\frac{\partial \bf {\vec f}}{\partial e } \;\frac{d \bf {\vec
\Phi}}{dt} \right]\;-\; \frac{2}{n\,a}\,\left[\frac{\partial
R}{\partial a }\,-\, {\bf {\vec \Phi}}\,\frac{\partial \bf {\vec
g}}{\partial a} \,-\,\frac{\partial  \bf {\vec f}}{\partial a }
\,\frac{d \bf {\vec \Phi}}{dt} \right]\;\;.
 \label{4.20}
 \ea
These gauge-invariant equations reveal the potential possibility
of simplification of orbit integration. One can deliberately
choose gauges different from (\ref{3.15}). In principle, it is
possible to pick up the gauge so as to nullify the right-hand
sides in three out of six equations (\ref{4.15} - \ref{4.20}).
This possibility is worth probing (we know from electrodynamics
that a clever choice of gauge considerably simplifies solution of
the equations of motion). Slabinski (2003) recently offered an
example of gauge that nullifies the rhs of equation (\ref{4.18}).

Another tempting possibility may be to pick up the gauge so that the
$\Phibold$-terms in (\ref{4.15} - \ref{4.20}) fully compensate the
short-period terms of the disturbing functions, leaving only the secular and
resonant ones. However, the afore discussed Newman's example strongly speaks
against this possibility, at least in the case of bound orbits. The case of
flybys seems to be more favourable, because in that case we do not have two
different time scales. Hence in that case a choice of some nonvanishing $
\Phibold$ may, potentially, lead to an even larger simplification of
calculations. We shall address this matter in a separate work.

Finally, we would point out that, in case the gauge $\;\Phibold\;$
depends not only upon time but also upon the "constants"
$\;C_i\;$, the right-hand sides of (\ref{4.15}) - (\ref{4.20})
implicitly contain time derivatives of $\;C_i\;$. Then, in order
to continue with analytic developments, it is necessary to
transfer such terms to the left-hand side, as we did in
(\ref{general_F}). This alteration will be dwelled upon in
(Efroimsky \& Goldreich 2003a,b).


\section{Delaunay's variables}

As well known (Brouwer \& Clemence 1961, p. 290), it is possible
to choose as the parameters $\;C_i\;$ not the six Keplerian
elements $\;C_i\;=\;\left(\,e
\,,\;a\,,\;M_o\,,\;\omega\,,\;\Omega\,,\;\inc\,\right)\;$ but the
set
$\;{D}_i\;=\;\left(\,L\,,\;G\,,\;H\,,\;M_o\,,\;\omega\,,\;\Omega\,\right)
\;$, new variables $\;L\;$, $\;G\;$, and $\;H\;$ being defined as
\ba L\,\equiv\,\mu^{1/2}\,a^{1/2}\,\;\;,\;\;\;\;\;
G\,\equiv\,\mu^{1/2}\,a^{1/2}\,\left(1\,-\,e^2\right)^{1/2}\,\;\;,\;\;\;\;
H\,\equiv\,\mu^{1/2}\,a^{1/2}\,\left(1\,-\,e^2\right)^{1/2}\,\cos
\inc\,\;\;\;, \label{6.1} \ea where
$\;\mu\;\equiv\;G(m_{\tiny{sun}}\,+\,m_{\tiny{planet}})\;$.

The advantage of these, Delaunay, variables lies in the
diagonality of their Lagrange-bracket matrix. Inversion thereof
yields, similarly to (\ref{4.15}) - (\ref{4.20}), the so-called
Delaunay system:
 \ba
\nonumber \frac{dL}{dt}\;\;=\;\;\frac{\partial R}{\partial M_o
}\;\;\;\;,\;\;\;\;\;\;
\frac{d M_o}{dt}\;=\;-\;\frac{\partial R}{\partial L}\;\;\;\;,\;\;\;\;\\
\frac{dG}{dt}\;\;=\;\;\frac{\partial R}{\partial \omega}\;\;\;\;\;\;,\;\;\;\;\;\;\;
\frac{d \omega}{dt}\;=\;-\;\frac{\partial R}{\partial G}\;\;\;\;,\;\;\;\;\;
 \label{6.2}\\
\nonumber \frac{d H}{dt}\;\;=\;\;\frac{\partial R}{\partial
\Omega}\;\;\;\;\;\;,\;\;\;\;\;\;\; \frac{d
\Omega}{dt}\;=\;-\;\frac{\partial R}{\partial
H}\;\;\;\;,\;\;\;\;\;
 \ea
{\underline{provided these parameters obey the Lagrange-type gauge
condition analogous to (\ref{3.9})}}:
 \be
\sum_i \;\frac{\partial {\tilde{\bf \vec f}}}{\partial
{D}_i}\;\frac{d {D}_i}{d t}\;=\;0\;\;.\;
 \label{6.3}
 \ee
where, similarly to (\ref{3} - \ref{4}),
 \be
\erbold\;=\;\tilde{\bf\vec
f}\left({D}_{1,...,6},\,t\right)\;\;\;,\;\;\; \;\;\;\;\;\;\;{\bf
\dot{\erbold}} \;=\;\tilde{\bf \vec g}\left( {D}_{1,...,6}, \,t
\right) \;\;\;\;\;\;,\;\;\;\;\;\;\frac{\partial \tilde{\bf \vec
f}}{\partial t}\;=\;\tilde{\bf \vec g}\;\;\;,
 \label{6.4}
 \ee
 and the tilde symbol emphasises that the functional dependencies of the position and
 velocity upon $\;D_i\;$ differ from their dependencies upon $\;C_i\;$.
We, thus, must keep in mind that the system of equations for the
Delaunay elements only pretends to exist in a 6-dimensional phase
space. In reality, it lives on a 9-dimensional submanifold
(\ref{6.3}) of a 12-dimensional manifold spanned by the Delaunay
elements and their time derivatives. In the case of analytical
calculations this, of course, makes no difference. Not in the case
of numerical computation, though.

If instead of the gauge condition (\ref{6.3}) we impose some
alternative gauge
 \be
 \sum_i \;\frac{\partial {\tilde{\bf \vec f}}}{\partial
{D}_i}\;\frac{d {D}_i}{d t}\;=\; \tilde{\bf \vec \Phi}
\left(t,\,{D}_{1,...,6} \right) \;\;\;,\;\;\;
 \ee
the generalised Delaunay-type equations will
read:\footnote{~Similarly to our comment in the end of Section 4
we would mention that if the gauge $\;\tilde{\Phibold}\;$ depends
not only on time but also on the parameters $\;C_i\;$ then the
right-hand sides of (\ref{6.5}) contain time derivatives of
$\;C_i\;$. A further analytic treatment of (\ref{6.5}) will then
demand that we transfer those terms to the left-hand side, as in
(\ref{general_F}). This will be discussed in (Efroimsky \&
Goldreich 2003a,b).}
 \ba
  \nonumber
\frac{dL}{dt}\,=\,\frac{\partial R}{\partial
M_o}\,-\,{\tilde{\bf\vec\Phi}} \frac{\partial \tilde{\bf \vec
g}}{\partial M_o}\,-\,\frac{\partial \tilde{\bf \vec f}}{
\partial M_o}\,\frac{d \tilde{\bf\vec\Phi} }{dt}\;\;\;,\;\;\;\;\;
\frac{d M_o}{dt}\;=\;-\;\frac{\partial R}{\partial
L}\;+\;{\tilde{\bf\vec\Phi}} \frac{\partial \tilde{\bf \vec
g}}{\partial  L}\;+\;\frac{\partial \tilde{\bf
\vec f}}{\partial L}\;\frac{d \tilde{\bf\vec\Phi} }{dt}\;\;\;,\;\;\;\;\;\\
\frac{dG}{dt}\;\;=\;\;\frac{\partial R}{\partial
\omega}\;-\;{\tilde{\bf\vec \Phi}}\frac{\partial \tilde{\bf \vec
g}}{\partial \omega} \;-\;\frac{\partial \tilde{\bf \vec
f}}{\partial \omega}\;\frac{d \tilde{\bf\vec\Phi} }{dt}\;\;\;\;
,\;\;\;\;\;\;\;\;\frac{d\omega }{dt}\;=\;-\;\frac{\partial
R}{\partial G}\;+\;{ \tilde{\bf\vec\Phi}} \frac{\partial
\tilde{\bf \vec g} }{\partial G} \;+\;\frac{\partial \tilde{\bf
\vec f}}{\partial G}\;\frac{d \tilde{\bf\vec\Phi}
}{dt}\;\;\;,\;\;\;\;\;
 \label{6.5}\\
\nonumber \frac{d H}{dt}\;\;=\;\;\frac{\partial R}{\partial
\Omega}\;-\;\tilde{\bf\vec \Phi}\frac{\partial \tilde{\bf \vec
g}}{\partial \Omega } \;-\;\frac{\partial \tilde{\bf\vec
f}}{\partial \Omega }\;\frac{d \tilde{\bf\vec\Phi}}{dt}\;\;\;,\;
\;\;\;\;\;\;\frac{d \Omega}{dt}\;=\;-\;\frac{\partial R}{\partial
H}\;+\;\tilde {\bf\vec\Phi}\frac{\partial \tilde{\bf\vec
g}}{\partial H}\;+\;\frac{\partial \tilde{\bf \vec f}}{\partial
H}\;\frac{d \tilde{\bf\vec\Phi}}{dt}\;\;\;,\;\;\; \;\;
 \ea
and the $\bf {\vec \Phi}$ terms should not be ignored, because
they account for the trajectory's deviation from the submanifold
(\ref{6.3}) of the ambient 12-dimensional space
$\;\left({D}_{1,...,6} \,,\;\dot{D}_{1,...,6}\right)\;$.

 The meaning of $\;\tilde{\bf\vec f}\;$
and $\;\tilde{\bf\vec g}\;$ in the above formulae is different
than that of $\;{\bf\vec f}\;$ and $\;{ \bf\vec g}\;$ in Sections
3 and 4. In those sections $\;{\bf\vec f}\;$ and $\;{\bf\vec g}\;$
denoted the functional dependencies (\ref{3}) of
$\;x\,,\;y\,,\;z\;$ and $\;\dot{x}\,,\;\dot{y}\,,\;\dot{z}\;$ upon
parameters $\;C_i\;=\;\left(\,e
\,,\;a\,,\;M_o\,,\;\omega\,,\;\Omega\,,\;\inc\,\right)\;$. Here,
$\;\tilde{\bf\vec f}\;$ and $\;\tilde{\bf\vec g}\;$ stand for the
dependencies of $\;x\,,\;y\,,\;z\;$ and
$\;\dot{x}\,,\;\dot{y}\,,\;\dot{z}\;$ upon the different set
$\;{D}_i\;=\;\left(\,L\,,\;G\,,\;H\,,\;M_o\,,\;\omega\,,\;
\Omega\,\right)\;$. Despite the different functional forms, the
{\it{values}} of $\;\tilde{\bf\vec f}\;$ and $\;\tilde{\bf\vec
g}\;$ coincide with those of $\;{\bf \vec f}\;$ and $\;{\bf\vec
g}\;$:
 \be
  \tilde{\bf\vec
f}\left({D}_{1,...,6}\right)\;=\;\erbold\;= \;{\bf\vec
f}\left({C}_{1,
...,6}\right)\;\;\;\;\;{\mbox{and}}\;\;\;\;\;\tilde{\bf\vec
g}\left({D}_{ 1,...,6}\right)\;=\;{\bf \dot{\erbold}} =\;{\bf\vec
g}\left({C}_{1,...,6}\right)\;\;\;.
 \label{6.6}
  \ee
Similarly, $\;\tilde{\bf \vec \Phi}\left({D}_{1,...,6}\right)\;$
and $\;{\bf \vec \Phi}\left({C}_{1,...,6}\right)\;$ are different
functional dependencies. It is, though, easy to show (using the
differentiation chain rule) that their values do coincide: \be
\tilde{\bf \vec \Phi}\left({D}_{1,...,6}\right)\;=\; {\bf \vec
\Phi}\left({C}_{1,...,6}\right)\;\; \label{6.7} \ee which is
analogous to the covariance of Lorentz gauge in electrodynamics.
This means that, for example, {\underline{analytical}}
calculations carried out by means of the Lagrange system
(\ref{4.6} - \ref{4.11}) are indeed equivalent to those performed
by means of the Delaunay system (\ref{6.2}), because imposure of
the Lagrange gauge $\;\bf \vec \Phi\;=\;0\;$ is equivalent to
imposure of $\;\tilde{\bf \vec \Phi}\;=\;0\;$.

Can one make a similar statement about numerical integrations?
This question is nontrivial. In order to tackle it, we should
recall that in the computer calculations the Lagrange condition
$\;\bf \vec \Phi\;=\;0\;$ cannot be imposed exactly, for the
numerical error will generate $\;some\;$ nonzero $\;\bf \vec
\Phi\;$. In other words, the orbit will never be perfectly
constrained to the submanifold $\;\bf \vec \Phi\;=\;0\;$. Thereby,
some nonzero $\;\bf \vec \Phi\; $ will, effectively, emerge in
(\ref{4.15} - \ref{4.20}). Similarly, a small nonzero
$\;\tilde{\bf \vec \Phi}\;$ will, effectively, appear in
(\ref{6.3}), and the Delaunay system will no longer be canonical.
In other words, we get not just an error in integration of the
canonical system, but we get an error that drives the system of
equation away from canonicity. This effect is not new: it is well
known that not every numerical method preserves the Hamiltonian
structure. Therefore, the unavoidable emergence of a nonzero
numerical-error-caused $\;\tilde{\bf \vec \Phi}\;$ in the Delaunay
system may, potentially, be a hazard. This topic needs further
investigation.

The gauge-invariant Delaunay-type system (\ref{6.5}) is no longer
Hamiltonian, unless the gauge is that of Lagrange,
$\;\Phibold\,=\,0\;$. It is possible to demonstrate that by
introducing a velocity-dependent disturbing force (which can be
done, for example, by defining the orbital elements in a
non-inertial reference frame) we further alter the Lagrange-type
and Delaunay-type gauge-invariant systems. The gauge-invariant
Delaunay-type system will remain non-canonical, except in one
special gauge which is called in (Efroimsky \& Goldreich 2003a)
"the generalised Lagrange gauge" and which simplifies to the
regular Lagrange gauge when the velocity dependence is removed.
Applications of this construction will be explained in Efroimsky
\& Goldreich (2003b).

To draw to a close, we would mention that in most books the
Lagrange and Delaunay systems of equations are derived not by a
straightforward application of the VOP method but through the
medium of Hamilton-Jacobi technique (Plummer 1918, Smart 1953,
Pollard 1966, Kovalevsky 1967, Stiefel and Scheifele 1971). At
first glance it may seem that such a derivation does not exploit
any sort of supplementary constraint and is, therefore, spared of
the gauge invariance. In our upcoming publication Efroimsky \&
Goldreich (2003a) we shall explain that the Lagrange constraint is
implicitly imposed within the Hamilton-Jacobi treatment of this
problem. Another method of derivation, different from both the VOP
and the Hamilton-Jacobi methods, was offered in the book by Kaula
(1968) and, once again, it may seem that his development is devoid
of whatever extra conditions. In the next section we explain
how Kaula's method, too, tacitly exploits the Lagrange constraint.\\

\section{Brouwer and Kaula}

As we saw above, the gauge-invariant generalisation of the
Delaunay system is no longer canonical. In celestial mechanics,
equations of type
 \ba
 \nonumber
\dot r\;=\;\frac{\partial H}{\partial p}\;+\;X(r,\,p)\\
 \label{59}\\
 \nonumber
\dot p\,=\,-\,\frac{\partial H}{\partial r}+P(r,\,p)
 \ea
were introduced by Dirk Brouwer who used them to include the
atmospheric-drag forces into the canonical picture. The quantities
$\;X\;$ and $\;P\;$ are called "canonical forces". The appropriate
formalism is comprehensively explained in Stiefel \& Scheifele
(1971). We see now that such "forces" can emerge not only due to
real physical interactions but also as purely mathematical
artifacts called into being by a nontrivial gauge choice. This in
no sense means that such terms may be omitted. Their relevance to
the stability of numerical methods will be discussed in Murison \&
Efroimsky (2003).

Above we saw that for a disturbance of type (\ref{4.110}) there
exists one special gauge (the Lagrange one) that makes the
equations for Delaunay's variables canonical. Still, in the case
of an arbitrary gauge the symplectic structure is
destroyed.\footnote{~It will be proven in our next paper
(Efroimsky \& Goldreich 2003a) that in the more general case of a
disturbing force dependent on positions and velocities the
situation is similar: the equations for the Delaunay elements will
become symplectic in a so-called generalised Lagrange gauge.} To
better understand the anatomy of this phenomenon, let us explore
an alternative derivation of Delaunay's equations, one based on a
direct change of variables. This method is akin to the VOP
technique, but has a better illustrative power. To simplify
things, we shall assume that the disturbing force depends on the
positions solely. Following Kaula (1968), we shall begin with the
canonical equations for the position and velocity in some inertial
Cartesian frame:
 \ba
 \nonumber
\frac{d \erbold }{dt}\,=\,\dot{ \erbold }\;\;\;,\\
 \label{60}\\
 \nonumber
\frac{d\,\dot{ \erbold }}{dt}\,=\,\frac{\partial U}{\partial
\erbold }\;\;.
 \ea
 A solution to these will depend upon seven quantities $\;S_0\equiv t,\,S_{1},\, ...\, , \,S_6\;$. One
 of these will be time, another six will be the parameters which
 can be devoid of or imparted with a time dependence of their own.
 This paves our way to the new, modified, system of equations:
 \ba
 \nonumber
\frac{\partial x_i }{\partial S_k}\,\frac{dS_k}{dt}=\,\dot{r_i }\;\;\;,\\
 \label{61}\\
 \nonumber
\frac{{ \partial {\dot r_i} }}{\partial
S_k}\,\frac{dS_k}{dt}\,=\,\frac{\partial U}{\partial
{{{r}_i}}}\;\;,
 \ea
summation over repeating indices, from $\;0\;$ through $\;6\,$,
being implied. After we multiply the upper and the lower equations
by $\;-\,\partial {\dot r}_i /
\partial S_{\it l}\;$ and $\;\partial {r}_i / \partial S_{\it
l}\;$, correspondingly, we should sum up both lines and arrive to:
 \ba
\left[S_{\it l}\,,\,S_k
\right]\;\frac{dS_k}{dt}\;=\;\frac{\partial }{\partial S_{\it l}}
\left(U\,-\,\frac{1}{2} {\dot r}_i {\dot r}_i
\right)\;\;\,,\;\;\;\;\;\;\;\left[S_{\it l}\,,\,S_k
\right]\;\equiv\;\frac{\partial r_i }{\partial S_{\it
l}}\;\frac{\partial \dot{r}_i }{\partial S_k}\;-\;\frac{\partial
\dot{r}_i }{\partial S_{\it l}}\;\frac{\partial r_i }{\partial
S_k}\;\;\;.
 \label{62}
 \ea
What remains now is to calculate the Lagrange-bracket matrix
$\;[S_{\it l}\,,\,S_k]\;$ and to invert the result. This will
yield the desired expressions for $\;dS_k/dt\;$. The expression
for $\;k\,=\,0\;$ will be merely an identity $\;1\,=\,1\;$,
because $\;S_0\equiv t\;$. The other six will be the planetary
equations. In case the parameters $\;S_1,\,...\,,\,S_6\;$ make a
Delaunay set, these equations will be expected to have a
Hamiltonian form. Have they really? To check this, one has to
calculate the appropriate Lagrange brackets. Or to accurately
examine how these are calculated in the literature. The most
direct way of computing the brackets is to differentiate formulae
(\ref{7}) with respect to the orbital elements. This will give the
well-known time-independent expressions for the Lagrange brackets.
These often-quoted standard expressions will, though, be valid in
the two-body case solely. The latter circumstance is missing in
(Kaula 1968) where the author implies that all will work equally
well in the case of N bodies. In reality, two types of additional
inputs will show themselves in the above formula for
$\;\bf{\dot{\vec{r}}}\;$ in the N-body problem: First of all,
extra items will appear in the expression for $\bf{\dot{{\vec
q}}}$:
 \ba
{\bf{\dot{\vec q}}}\;=\;\frac{\partial {\bf{\vec q}}}{\partial t}
\;+\;\frac{\partial {\bf{\vec q}}}{\partial a} \frac{d {{a
  }}}{dt}
\;+\;\frac{\partial {\bf{\vec q}}}{\partial e} \frac{d {{e
  }}}{dt}
\;+\;\frac{\partial {\bf{\vec q}}}{\partial M_o} \frac{d {{M_o
  }}}{dt}\;\;.
 \label{66}
 \ea
Beside this, the contribution $\;{\bf{\dot R}}{\bf{\vec q}}\;$
will now enter the expression for $\;\bf{\dot{\vec{x}}}\;$. Here
 \ba
{\bf{\dot R}}\;=\;\frac{\partial {\bf{R}}}{\partial \Omega}
\frac{d {{\Omega }}}{dt}\;+\;\frac{\partial {\bf{R}}}{\partial
\omega} \frac{d {{\omega  }}}{dt}\;+\; \frac{\partial
{\bf{R}}}{\partial \inc} \frac{d {{\inc
  }}}{dt}\;\;.
 \label{67}
 \ea
Altogether, these will result in the following expression for the
velocity:
 \ba
 \nonumber
{\bf {\dot{\erbold}}}\;=\;{\bf R}(\Omega,\,\inc,\,\omega)\;{\bf
{\dot{\vec q}}}(a,\,e,\,M)\;+\;{\bf{\dot R}}{\bf{\vec q}}
 \ea
 \ba
 \nonumber
  =\;{\bf R}\;\frac{\partial{\bf {{\vec q}}}}{\partial
t}\;+\;\frac{\partial ({\bf R}{\bf{\vec q}})}{\partial a} \frac{d
{{a
  }}}{dt}
\;+\;\frac{\partial ({\bf R}{\bf{\vec q}})}{\partial e} \frac{d
{{e
  }}}{dt}
\;+\;\frac{\partial ({\bf R}{\bf{\vec q}})}{\partial M_o} \frac{d
{{M_o
  }}}{dt}
\;+\;\frac{\partial ({\bf{R}} {\bf{\vec q}})}{\partial \Omega}
\frac{d {{\Omega }}}{dt}\;+\;\frac{\partial ({\bf{R}}{\bf{\vec
q}})}{\partial \omega} \frac{d {{\omega  }}}{dt}\;+\;
\frac{\partial ({\bf{R}}{\bf{\vec q}})}{\partial \inc} \frac{d
{{\inc }}}{dt}
 \ea
 \ba
 =\;{\bf R}\;\frac{\partial{\bf {{\vec q}}}}{\partial
t}\;+\;{\bf{\vec \Phi}} \;\;\;.
 \label{68}
 \ea
In Kaula (1968) the latter term is ignored, so that the author
implicitly imposes the Lagrange gauge. In this gauge,
differentiation of the above expression with respect to the
elements will indeed entail the standard time-independent
expressions for the Lagrange brackets.  In a non-Lagrange gauge,
though, the situation will be different, and we shall not end up
with a canonical Delaunay system. Instead, we shall arrive to the
non-canonical gauge-invariant Delaunay-type system (\ref{6.5}).

Very similarly, the Lagrange constraint enters all the methods by
which the Gauss system of equations is derived in the literature,
and the imposure of this constraint is camouflaged in a manner
similar to what we saw above in regard to the Delaunay equations.
We shall not engage in a comprehensive discussion of this issue,
but shall rather provide a typical example. In section 11.5 of
(Danby 1988) an auxiliary vector $\;\bf{\hat h}\;$ is introduced
as a unit vector aimed along the instantaneous orbital momentum of
the body, relative to the gravitating centre. Then $\;\bf{\hat
h}\;$ gets resolved along the inertial axes $\;(x,\;,y,\;z)\;$,
where $\;{\bf{\hat x}}\;$ points toward the vernal equinox and
$\;{\bf{\hat z}}\;$ toward the north of the ecliptic pole:
 \ba
 \nonumber
 {\bf{\hat h}\;}=\;{\bf{\hat x}}\;\sin \Omega\;\sin \inc\;
                -\;{\bf{\hat y}}\;\cos \Omega\;\sin \inc\;
                +\;{\bf{\hat z}}\;\cos \inc\;\;.
 \ea
This expression is certainly correct in the two-body case. It
remains valid also in the N-body problem, {\it{only if the orbital
elements $\;\Omega\;$ and $\;\inc\;$ are osculating}}, i.e., only
if the instantaneous inertial velocity is tangential to the
ellipse parametrised by the Keplerian set that includes these
$\;\Omega\;$ and $\;\inc\;$. Thus, the Lagrange constraint is
implied.\\

\section{Is it worth it?}

At this point one may ask if it is at all worth taking the nonzero
$\;\tilde{\bf \vec \Phi}\;$ into account in the Delaunay
equations. After all, one can simply restrict himself to the
6-dimensional phase space defined by $\;{D}_i\;$, and
{\it{postulate}} that the six unwanted extra dimensions
$\;\dot{{D}}_i\;$ do not exist (i.e., postulate that $\;\tilde{\bf
{\vec \Phi}}\;=\;0\;$). This, of course, can be done, but only at
some cost: a certain type of accumulating numerical error will be
ignored (not eliminated), and it will keep accumulating. As
explained in the end of the previous section, the overall
integration error of a Hamiltonian system consists of an error
that leaves the system canonical (like, for example, an error in
calculation of $\;R\;$ in (\ref{6.5})) and an error that drives
the system away from its canonicity (like the error reflected in
the accumulated nonzero value of $\,{\bf {\vec \Phi}})\,$. The
$\,{\bf \vec \Phi}\,$ terms in (\ref{6.5}) play the role of
correctors: they fully compensate for the errors of the second
type (i.e., for what in numerical electrodynamics is called gauge
shift).

Similarly, in the case of Lagrange system, one may enquire if it
is worth introducing the 12-dimensional space spanned by the
orbital elements $\,C_i$ and their time derivatives $\,H_i
\equiv\,\dot{C}_i$. Why not simply consider a trajectory in the
6-dimensional space of $\,C_i$ and assume that the Lagrange gauge
is fixed exactly? Indeed, if we are solving the problem
$\,{\bf{\ddot {\vec r}}}\,=\,{\bf {\vec f}}({\bf{\vec r}})\,$, is
it worth introducing an extra entity $\,{\bf
{\vec{v}}}\,\equiv\,{\bf{\dot{\vec r}}}\,$ and considering the
orbit integration in the space of a larger dimension, spanned by
the components of both $\,\bf {\vec r}\,$ and $\,\bf {\vec v}\,$?
Will this new entity $\,\bf {\vec v}\,$ add anything? The answer
to this question will be affirmative if we take into account the
fact that a trajectory is not merely a locus of points visited by
the body: the notion of trajectory also comprises the {\it rate}
at which the body was travelling. Appropriately, the accumulated
numerical error will consist of two parts: distortions of the
orbit shape, and distortions in the time-dependence of the speed
at which the orbit was described. This explains why the events are
taking place not just in the space of orbital elements but in the
larger space of the elements and their time derivatives. This
issue is best of all explained by Hagihara (1970). In subsection
1.6 of the first volume of his treatise he contrasts the cases of
exact and apparent equivalences of dynamical systems. (The latter
case is that of the orbital curves being geometrically, not
dynamically, identical.)

Still, there is more to it, because a convenient choice of gauge
may simplify the solution of the equations of motion (Slabinski
2003; ~Efroimsky \& Goldreich 2003b).


\section{Conclusions}

We have demonstrated a previously unrecognised aspect of the
Lagrange and Delaunay systems of planetary equations. Due to the
Lagrange gauge condition (\ref{3.9}), the motion is, in both
cases, constrained to a 9-dimensional submanifold of the ambient
12-dimensional space spanned by the orbital elements and their
time derivatives. Similarly to the field theory, the choice of
gauge is vastly ambiguous and reveals a hidden symmetry (and an
appropriate symmetry group) inherent in the description of the
N-body problem in terms of the orbital elements. Just as a choice
of a particular gauge simplifies solution of the equations of
motion in electrodynamics, an alternative (to that of Lagrange)
choice of gauge can simplify orbit calculations. We have written
down the generalised Lagrange-type (\ref{4.15} - \ref{4.20}) and
Delaunay-type (\ref{6.5}) equations in their gauge-invariant, form
(with no specific gauge imposed). We have pointed out that neither
the Lagrange gauge (\ref{3.9}) nor any other constraint is exactly
preserved in the course of numerical computation. This may be a
source of numerical
instability.\\


{\bf Acknowledgements}\\
~\\
Debts of gratitude are always pleasure to pay. I am grateful to
William Newman for our numerous conversations on planetary
dynamics. I also wish to acknowledge the critical questions and
comments by Yurii Batrakov, Robert Burridge, Sergei Klioner, Mark
Levi, Robert O'Malley, Scot Tremaine, and Jack Wisdom, which
greatly helped me to improve the paper. My truly special thanks go
to Peter Goldreich who kindly discussed with me some relevant
concepts, and offered very important corrections and amendments to
the manuscript. This research was supported by NASA
grant W-19948.\\


\end{document}